\documentclass[entropy,article,moreauthors,pdftex,10pt]{mdpi} 

\firstpage{1} 
\articlenumber{x}
\doinum{10.3390/------}
\pubvolume{xx}
\pubyear{2017}
\copyrightyear{2017}
\externaleditor{Academic Editor:  Mariela Portesi, Alejandro Hnilo, and Federico Holik}
\history{Received: date; Accepted: date; Published: date}
\pdfoutput=1

\usepackage{amsmath}
\usepackage{amsfonts}
\usepackage{stmaryrd}

\preto{\abstractkeywords}{\nolinenumbers}

 \theoremstyle{mdpi}
 \newcounter{thm}
 \newcounter{ex}
 \newcounter{re}
 \theoremstyle{mdpidefinition}

 \newtheorem{Definition}[thm]{Definition}

\Title{Contextuality and Indistinguishability}

\Author{J. Acacio de Barros $^{1,}$*, Federico Holik $^{2}$ and D\'{e}cio Krause $^{3}$}
\AuthorNames{J. Acacio de Barros, Federico Holik and D\'{e}cio Krause}

\address{%
$^{1}$ \quad School of Humanities and Liberal Studies, San Francisco State University, San Francisco, CA, USA; barros@sfsu.edu\\
$^{2}$ \quad Instituto de F\'{\i}sica La Plata, UNLP, CONICET, Facultad de Ciencias Exactas, C.C.~67, 1900 La Plata, Argentina; olentiev2@gmail.com\\
$^{3}$ \quad Department of Philosophy, Federal University of Santa Catarina, Florian\'{o}polis, SC, Brazil; deciokrause@gmail.com}

\corres{Correspondence: barros@sfsu.edu}

\firstnote{All authors contributed equally to this work.}

\abstract{It is well known that in quantum mechanics we cannot always define consistently properties that are context independent. Many approaches exist to describe contextual properties, such as Contextuality by Default (CbD), sheaf theory, topos theory, and non-standard or signed probabilities. In this paper we propose a treatment of contextual properties that is specific to quantum mechanics, as it relies on the relationship between contextuality and indistinguishability. In particular, we propose that if we assume the ontological thesis that quantum particles or properties can be indistinguishable yet different, no contradiction arising from a Kochen-Specker-type argument appears: when we repeat an experiment, we are in reality performing an experiment measuring a property that is indistinguishable from the first, but not the same. We will discuss how the consequences of this move may help us understand quantum contextuality.}

\begin{document}

\section{Introduction}

Quantum mechanics does not allow for the simultaneous measurement of complementary properties. This is exemplified by the famous case of momentum and position: the experimental setups required to measure them are incompatible, which means that they cannot be measured together. This fact is expressed in the commutation relation $[\hat{x}, \hat{p}] = i \hbar$, where  $\hat{x}$  is the position and $\hat{p}$ the momentum operators. Non-commuting operators do not share all their eigenvectors, and it is possible to find a quantum state that has a sharply defined value for, say, position (e.g., $\delta (x)$, where $\delta$ is the Dirac function\footnote{Not really a function, but a distribution---see \cite{antoine_rigged_2009}}), but whose complementary property is not sharply defined (in the case of $\delta(x)$, the momentum can be anywhere between $-\infty$ and $\infty$).  So, for complimentary properties it seems that quantum mechanics forbids us from prescribing them well-defined values. 

But is it really true that if properties cannot be measured simultaneously then it is impossible to assign simultaneous values to them? This was, in fact, the question behind the argument put forth by Einstein, Podolsky, and Rosen (EPR) \cite{einstein_can_1935} in their famous 1935 paper. In it, EPR argued that for special two-particle entangled systems, one could know the value of a property in one of its particles without actually performing an experiment on it, due to correlations encoded in the entangled wave-function. Therefore, for two complementary properties, e.g. momentum and position, though one could not measure them simultaneously,  one could  assign values to them. EPR then argued that the description of nature based on the wave-function, which did not include simultaneous values of complementary properties, was incomplete. These more "complete" theories, ones that could describe the properties of quantum systems from unobservable hidden states, became known as \textit{hidden-variable theories}, as they used \textit{hidden variables} that themselves would not be directly observable. 

The debate about the existence of hidden variables was intense, and giving a historical account of it would go beyond the scope of this paper. However, we want to point out a couple of landmark results that challenged this research program. As early 1932, Wigner showed that a joint probability distribution for two complimentary properties, in this case  momentum and position, consistent with quantum statistical mechanics had to have negative values, therefore not being a proper probability distribution \cite{wigner_quantum_1932}. This result suggested that an attempt to simultaneously define momentum and position had at least some serious technical challenges. Later on, in his famous book on the mathematical structure of quantum mechanics, von Neumann proved a no-go theorem for hidden-variable theories which discouraged many of pursuing them. However, several decades later, John Bell realized that von Neumann's assumptions were too strong \cite{bell_speakable_2004}, and that the no-go theorem was, in Bell's own words, ``silly.'' So silly in fact that, before Bell, in 1952 David Bohm \cite{bohm_suggested_1952,bohm_suggested_1952-1} had already created a hidden-variable theory that accounted for all the experimental outcomes of quantum mechanics, thus ``disproving'' von Neumann's no-go theorem.

But the main results challenging the concept of well-defined properties for quantum systems came with the theorems of Bell and of Kochen and Specker. Bell showed that locality and well-defined values of a quantity before a measurement (realism) was inconsistent with the predictions of quantum mechanics \cite{bell_einstein-podolsky-rosen_1964}. But perhaps more relevant was a no-go theorem by Kochen and Specker (KS) \cite{kochen_problem_1967}. KS showed that for a Hilbert space  $\mathcal{H}$ of dimension greater than two, it is possible to construct a set of True/False properties (projection operators in $\mathcal{H}$) that commute in a given context (i.e., that can be simultaneously measured), but that no truth value can be globally assigned to them in the totality of contexts. The main reason is that the truth value of a property needs to change if we observe it together with one set of other properties or with another set (context). This is the idea of \textit{contextuality}: properties change (in this case, their truth values) from one context to another. KS  proved that quantum observables (properties) are contextual. 

Intuitively, if a property changes from one context to another, this presents a problem, if we think about properties in terms of the standard setting, e.g. in terms of classical predicate logic. For example, in classical predicate logic the system $S$ has property $P$ if the proposition $P(S )$ has truth value ``true.'' However, for the KS set of quantum observables, it is not possible to assign a truth value to $P(S)$ in a consistent way for all contexts.  Physicists assume all occurrences of $P$ as being \textit{the same} $P$, as if every time we were to measure a certain observable, we would measure \textit{the same} property. This assumption carries with it a very strong ontological conjecture, as we will show below. Indeed, we will explore the possibility that the standard theory of identity does not applies, and thus, property $P$ cannot be discerned in the distinct contexts. From this perspective, the difficulty in defining properties in quantum mechanics would originate in the fact that we cannot, from the Hilbert space formalism or even the experimental setup, apply the standard theory of identity to properties and particles in different contexts.

Thus, in this work we present a new look at how to consider properties, that is, to consider a formal theory of properties and entities in which they can be seen as not being \textit{the same}, but still being \textit{indistinguishable} in different contexts. Our proposed theory of indistinguishable properties would be slightly different from simply saying that properties are context-dependent, an approach partially espoused by Dzhafarov and Kujala in \cite{dzhafarov_contextuality-by-default_2016} (although these authors have not moved to "non-classical  ontological" settings, as we do). We will represent properties of particles by \textit{indistinguishable} predicates in different contexts.

Our main idea runs as follows. Using \textit{quasi-set theory}, a mathematical theory where we can deal with indistinguishable but not identical objects (something we cannot rightly do in standard mathematics, as we will see with more details at section \ref{sec:qsets}), we can define \textit{indistinguishable properties}. The intuition is that we neither  perform ``the same'' experiment twice, nor measure two indistinguishable  properties  on ``a same'' quantum system either, but we measure indistinguishable properties (prepared the ``same''  way) over indistinguishable quantum systems. In other words, we need to seriously consider  the notion of  indistinguishability (or indiscernibility) as something distinct from identity (as we shall see, these notions are confounded in standard logic and mathematics). Then, with these concepts at hands, we can read again the results  by KS and realize that the core of  their theorem (the "paradox") can be avoided, for the contradiction assumes that ``the same'' properties are measured in ``the same'' particles in different contexts. But, if we realize that we measure \textit{indistinguishable} properties over \textit{indistinguishable} particles, there will be no surprise in acknowledging that the obtained results may differ. The problem, as we intend to develop in this paper in a rough but yet mathematically precise form, is to provide a formalism for defining or considering \textit{legitimate} (and not \textit{fake}) indiscernible objects and properties.

Our paper is organized the following way. In Section \ref{sec:KS-argument} we detail the contextuality argument used by KS in terms of probability spaces, in order to generalize it to more realistic situations where we do not need probability-one events.  We then re-think the KS concept of contextuality in terms of its implication to the concept of distinguishability, and show that  an essential component of KS's proof is that the properties they used are assumed to obey the classical theory of identity.  Given the motivation for thinking about indistinguishability in connection to quantum contextuality, presented in \ref{sec:Quasi-Properties}, in Section \ref{sec:ontology} we discuss how we can implement such concepts in a precise way, both ontologically and mathematically (as in \ref{sec:qsets}).  Finally, in Section \ref{sec:indisting-assumption} we show explicitly the connection between quantum contextuality and indistinguishability, by constructing an explicit concept  of indistinguishable property that does not lead to a KS-type contradiction. We end in Section  \ref{sec:Conclusions} with some conclusions and possible open questions. 

\section{KS argument for contextuality}\label{sec:KS-argument}

Let us begin by stating some basic concepts that will help us connect the issue of quantum properties with indistinguishability. Let us start by defining  contextuality as it is relevant for quantum mechanics: from the structure of a probability space. Since the KS theorem works for Hilbert spaces of dimension greater than two, it is not necessary, for our purposes, to deal with the mathematical difficulties originated by using infinite sample spaces, so here we use only finite sets.  Komogorov \cite{kolmogorov_foundations_1956} defined probabilities in an axiomatic way as follows.
\begin{Definition}[Kolmogorov]\label{def:probability}
The triple $\mathfrak{P} = \left(\Omega,\mathcal{F},p\right)$ is a probability space if $\Omega$ is a finite set (the \emph{sample space}), $\mathcal{F}$ is an algebra over $\Omega$, and $p:\mathcal{F}\rightarrow\left[0,1\right]$ is a function satisfying the following axioms:
\begin{description}
\item [{K1}] $p\left(\Omega\right)=1$
\item [{K2}] $p\left(A\cup B\right)=p\left(A\right)+p\left(B\right)$, for all $A$ and $B$ in $\mathcal{F}$ such that $A\cap B=\emptyset$.
\end{description}
\end{Definition}
We represent the outcomes of experiments in terms of random variables, which are measurable functions that take numerical values corresponding to such outcomes. 
\begin{Definition}\label{def:random-variable}
Let $\mathfrak{P} = \left(\Omega,\mathcal{F},p\right)$ be a probability space, and $S$ a finite set of real numbers (corresponding to possible experimental outcomes) and $\mathcal{T}$ an algebra over $S$. A random variable $\mathbf{A}$ in this probability space is a measurable function $\mathbf{A}:\mathcal{F}\rightarrow S$, i.e., a function such that for every $T\in\mathcal{T}$, $\mathbf{A}^{-1}\left(T\right)\in\mathcal{F}$.
\end{Definition}
Intuitively, each element of $\Omega$ is randomly selected with a probability given by $p$, and when a particular element is selected, the function $\mathbf{A}$ produces an outcome in $S$. The inverse of $\mathbf{A}$ produces a measurable partition in $\mathcal{F}$ corresponding to different values of possible experimental outcomes. When representing the outcomes of an experiment, a probability space and a random variable, with its corresponding partitions, must then be constructed such that the random variable has the same stochastic behavior as the observed experimental outcomes.  
\begin{Definition}\label{def:expectation}
The expectation of an $S$-valued random variable $\mathbf{A}$, $E\left(\mathbf{A}\right)$,
is 
\[
E\left(\mathbf{A}\right)=\sum_{s\in S}sp\left(\mathbf{A}=s\right).
\]
The expectation of the product an $S$-valued random variables $\mathbf{A}$ and an $S'$-valued random variable $\mathbf{B}$, also called their second moment, is
\[
E\left(\mathbf{A}\mathbf{B}\right)=\sum_{s\in S}\sum_{s'\in S'}ss'p\left(\mathbf{A}=s,\mathbf{B}=s'\right).
\]
The expectation of the product an $S$-valued random variables $\mathbf{A}$, an $S'$-valued random variable $\mathbf{B}$, and an $S''$-valued random variable $\mathbf{C}$,  called their third moment, is
\[
E\left(\mathbf{A}\mathbf{B}\mathbf{C}\right)=\sum_{s\in S}\sum_{s'\in S'}\sum_{s'\in S''}ss's''p\left(\mathbf{A}=s,\mathbf{B}=s',\mathbf{C}=s''\right).
\]
The fourth moment, fifth moment, etc. are defined in a similar way. The probabilities $p\left(\mathbf{A}=s,\mathbf{B}=s'\right)$, $p\left(\mathbf{A}=s,\mathbf{B}=s',\mathbf{C}=s''\right)$,$\ldots$,  $p\left(\mathbf{A}=s,\mathbf{B}=s',\ldots, \mathbf{Z}=s^{n}\right)$   are called the \emph{joint probability} for $\mathbf{A}$ and $\mathbf{B}$, $\mathbf{A}$, $\mathbf{B}$, and $\mathbf{C}$, etc. 
\end{Definition}

To understand what is contextuality, we now examine a simple example dating back to Boole, but related to the discussions in quantum mechanics by Specker's parable of the overprotective seer \cite{specker_logic_1975}. We start with a set of properties, $X$, $Y$, and $Z$, that can be either true or false for each running of an experiment about a certain system of interest. In the most general case of interest, such properties could be stochastic, and therefore we would need to represent them within the formalism of probability theory. To do so, let us consider a set of three $\pm1$-valued random variables\footnote{Random variables where $S=\left\{ 1,-1\right\} $.}, $\mathbf{X}$, $\mathbf{Y}$, and $\mathbf{Z}$, with ``$+1$'' corresponding to the property being ``true'' and ``$-1$'' to ``false''. Let us further assume that, experimentally, our constraint is that we cannot observe the properties $X$, $Y$, and $Z$ simultaneously, but we can only observe them one at a time or in pairs\footnote{An explicit example using a firefly in a box is provided in \cite{de_barros_negative_2016}.}. Suppes and Zanotti \cite{suppes_when_1981} showed that in such case, there exists a probability space, with a corresponding joint probability distribution, for $\mathbf{X}$, $\mathbf{Y}$, and $\mathbf{Z}$ if and only if 
\begin{eqnarray}
-1 & \leq & E\left(\mathbf{XY}\right)+E\left(\mathbf{XZ}\right)+E\left(\mathbf{YZ}\right)\label{eq:suppes-zanotti}\\
 & \leq & 1+2\min\left\{ E\left(\mathbf{XY}\right),E\left(\mathbf{XZ}\right),E\left(\mathbf{YZ}\right)\right\} .\nonumber 
\end{eqnarray}

What happens when (\ref{eq:suppes-zanotti}) is violated? To see this, let us consider the extreme case of maximum violation of the left hand side of (\ref{eq:suppes-zanotti}): $E\left(\mathbf{XY}\right)=E\left(\mathbf{XZ}\right)=E\left(\mathbf{YZ}\right)=-1$. It is easy to see that this is mathematically (and logically, if we think about truth values) impossible: if $\mathbf{X}=1$, then $E\left(\mathbf{XY}\right)=-1$ implies $\mathbf{Y}=-1$ with probability $1$, which from $E\left(\mathbf{YZ}\right)=-1$ we obtain $\mathbf{Z}=1$, and finally from $E\left(\mathbf{XZ}\right)=-1$ we get $\mathbf{X}=-1$, a clear contradiction. A contradiction is also obtained for $\mathbf{X}=-1$. 

The above contradiction may lead us to believe that (\ref{eq:suppes-zanotti}) can never be violated. However, this is not necessarily the case \cite{de_barros_quantum-like_2012,de_barros_decision_2014,de_barros_beyond_2015}, as the property $X$ is observed in two different experimental situations: (i) $X$ together with $Y$, and (ii) $X$ together with $Z$. Since the contexts (experimental conditions) are different, it is possible for the property $X$ to change from situation (i) to (ii). When this happens, we call the properties $X$, $Y$, and $Z$, or their corresponding random variables, $\mathbf{X}$, $\mathbf{Y}$, and $\mathbf{Z}$, \emph{contextual}. 
\begin{Definition}
\label{def:contextual}
Let $A=\left\{ A_{1},A_{2},\ldots,A_{n}\right\} $, $n\geq3$, be a collection of properties observable in a multitude of experimental conditions. This collection is \emph{non-contextual} if an only if there exists a probability space $\left(\Omega,\mathcal{F},p\right)$ and a collection of random variables, $\mathbf{A}=\left\{ \mathbf{A}_{1},\mathbf{A}_{2},\ldots,\mathbf{A}_{n}\right\} $, $\mathbf{A}_{i}:\mathcal{F}\rightarrow E_{i}$, on $\left(\Omega,\mathcal{F},p\right)$, such that all observable stochastic properties of $A$ are represented by $\mathbf{A}$. Otherwise, the collection of properties $A$ is \emph{contextual}. 
\end{Definition}
In other words, a collection of properties $\left\{ A_{1},A_{2},\ldots,A_{n}\right\} $, $n\geq3$, is contextual if and only if no joint probability distribution for all random variables $\mathbf{A}_{i}$ representing properties $A_{i}$ exist in a probability space $\left(\Omega,\mathcal{F},p\right)$. 

As we saw from the example and definitions above, properties are said to be contextual if we cannot create a single probability space that consistently represent those properties. To better understand this, let us connect Definition \ref{def:contextual}  to our three random-variable example discussed above.  Let us assume that we obtained the value $\mathbf{X}=1$ and  $\mathbf{Y}=-1$ in a given experiment. The existence of a probability space $\mathfrak{P}$ assures us that there is an element of $\mathcal{F}$, call  $f\in \mathcal{F}$, such that $\mathbf{X}(f)=1$ and  $\mathbf{Y}(f)=-1$. However, this very same element, when used in the random variable $\mathbf{Z}$ will give either $-1$ or $1$, which would not yield the inconsistent anti-correlations  $E\left(\mathbf{XY}\right)=E\left(\mathbf{XZ}\right)=E\left(\mathbf{YZ}\right)=-1$. On the other hand, were the anti-correlations  experimentally observed, it is clear that the $f\in \mathcal{F}$  used in one experimental context cannot be the same as in another experimental context, as this would result in consistent correlations. The very same argument can be used for any inconsistent correlations, such as  $E\left(\mathbf{XY}\right)=E\left(\mathbf{XZ}\right)=E\left(\mathbf{YZ}\right)=-1$, $E\left(\mathbf{XY}\right)=E\left(\mathbf{XZ}\right)=1=-E\left(\mathbf{YZ}\right)$, etc. In fact, it can be shown (see \cite{abramsky_logical_2012}) that any impossibility of obtaining a joint probability distribution amounts to some combination of logical inconsistencies, such as the correlations above. 

What is happening in contextual systems is that calling a property $A_{i}$ in a context the same as $A_{i}$ in a different context is a mistake, as it leads to inconsistencies. A clear approach to resolve those inconsistencies, one advocated by Dzhafarov and Kujala, is to label variables according to their context \cite{dzhafarov_contextuality-by-default_2016}. This approach is called Contextuality by Default (CbD). According to it, we would not have only property $A_{i}$, but instead, say, at least two different properties, $A_{i,1}$ and $A_{i,2}$, where $1$ and $2$ refer to different experimental conditions (of course, more experimental conditions would require more properties). Explicitly, in the $X$, $Y$, and $Z$ example, since we have three experimental conditions, the properties would be $X_{1}$, $X_{2}$, $Y_{1}$, $Y_{3}$, $Z_{2}$, and $Z_{3}$, and with this extended set or properties, no contradiction would appear. 

The three random-variable  example above is useful for us to understand the concept of contextuality, but it  is not an example that  comes from quantum mechanics. In fact, it is easy to show that for three quantum observables in a Hilbert space,  $\hat{X}$,  $\hat{Y}$,  and $\hat{Z}$,  with eigenvalues  $\pm 1$, if they pairwise commute, i.e.   $[\hat{X},\hat{Y}]=[\hat{X},\hat{Z}]=[\hat{Y},\hat{Z}]=0$, then  they are not contextual \cite{specker_logic_1975}.  Therefore, we cannot get the contextuality exemplified above from a physical quantum system. To provide a more physically grounded example, let us examine the famous Kochen-Specker (KS) theorem \cite{kochen_problem_1967}, in the simpler version with 18-vectors given by Cabello et al. \cite{cabello_bell-kochen-specker_1996}.  Here we use a four-dimensional  Hilbert space, and as  such, we can find groups of four orthogonal vectors whose corresponding projectors commute.  Consider, for instance, the non-normalized and orthogonal vectors $\vec{a}=(0,0,0,1)$, $\vec{b}=(0,0,1,0)$, $\vec{c}=(1,1,0,0)$, $\vec{d}=(1,-1,0,0)$. Their corresponding projectors can defined  as the matrix that projects any vector into the subspace spanned by them. For example, applying the projector  $\hat{P}_{0,0,0,1}$  associated to  $\vec{a}$ to the vector $(x_1,x_2,x_3,x_4)$ would yield the vector   $(0,0,0,x_4)$, whereas    $\hat{P}_{0,0,1,0}$  associated to  $\vec{b}$ yields    $(0,0,x_3,0)$, and so on.  Since  $\vec{a}$,  $\vec{b}$,  $\vec{c}$,  and $\vec{d}$ are orthogonal to each other, their projectors commute (e.g.  $[\hat{P}_{0,0,1,0},\hat{P}_{0,0,0,1}]=0$), which means that they correspond to observables that can be measured simultaneously.  Furthermore, since the Hilbert space is four dimensional, it also follows that 
\begin{equation}
\hat{P}_{0,0,1,0} + \hat{P}_{0,0,1,0} + \hat{P}_{1,1,0,0} + \hat{P}_{1,-1,0,0} = \hat{1},
\end{equation}
where $\hat{1}$  is the identity operator. 

In a four dimensional Hilbert space, let us consider now the following set of projectors: 
\begin{align}
\hat{P}_{0,0,0,1}+\hat{P}_{0,0,1,0}+\hat{P}_{1,1,0,0}+\hat{P}_{1,-1,0,0} & =\hat{1},\label{eq:cabello-1st}\\
\hat{P}_{0,0,0,1}+\hat{P}_{0,1,0,0}+\hat{P}_{1,0,1,0}+\hat{P}_{1,0,-1,0} & =\hat{1},\\
\hat{P}_{1,-1,1,-1}+\hat{P}_{1,-1,-1,1}+\hat{P}_{1,1,0,0}+\hat{P}_{0,0,1,1} & =\hat{1},\\
\hat{P}_{1,-1,1,-1}+\hat{P}_{1,1,1,1}+\hat{P}_{1,0,-1,0}+\hat{P}_{0,1,0,-1} & =\hat{1},\\
\hat{P}_{0,0,1,0}+\hat{P}_{0,1,0,0}+\hat{P}_{1,0,0,1}+\hat{P}_{1,0,0,-1} & =\hat{1},\\
\hat{P}_{1,-1,-1,1}+\hat{P}_{1,1,1,1}+\hat{P}_{1,0,0,-1}+\hat{P}_{0,1,-1,0} & =\hat{1},\\
\hat{P}_{1,1,-1,1}+\hat{P}_{1,1,1,-1}+\hat{P}_{1,-1,0,0}+\hat{P}_{0,0,1,1} & =\hat{1},\\
\hat{P}_{1,1,-1,1}+\hat{P}_{-1,1,1,1}+\hat{P}_{1,0,1,0}+\hat{P}_{0,1,0,-1} & =\hat{1},\\
\hat{P}_{1,1,1,-1}+\hat{P}_{-1,1,1,1}+\hat{P}_{1,0,0,1}+\hat{P}_{0,1,-1,0} & =\hat{1},\label{eq:cabello-last}
\end{align}
where $\hat{1}$  is the identity operator in this space. Each line in the set of equations above has four commuting observables to whose outcomes we can attribute truth values. The fact that each line sums to one simply states that one, and only one, property per line is true. 

The KS contradiction is obtained by assuming a sample space $\Omega$ and realizing that for an $\omega\in\Omega$ it is not possible to assign values for the properties associated to the projectors in a consistent way. To see this, let us rewrite the projection operator equations in terms of outcomes of experiments, i.e. using random variables $\mathbf{P}_{i}$'s taking values 0 or 1 (for ``false'' and ``true,'' respectively).  These random variables would correspond to each line in (\ref{eq:cabello-1st})--(\ref{eq:cabello-last}), and would depend explicitly on the element $\omega$ of the sample space. This leads to 
\begin{align}
\mathbf{P}_{0,0,0,1}\left(\omega\right)+\mathbf{P}_{0,0,1,0}\left(\omega\right)+\mathbf{P}_{1,1,0,0}\left(\omega\right)+\mathbf{P}_{1,-1,0,0}\left(\omega\right) & =1,\label{eq:cabello-1st-omega}\\
\mathbf{P}_{0,0,0,1}\left(\omega\right)+\mathbf{P}_{0,1,0,0}\left(\omega\right)+\mathbf{P}_{1,0,1,0}\left(\omega\right)+\mathbf{P}_{1,0,-1,0}\left(\omega\right) & =1, \label{eq:KS-omega-2}\\
\mathbf{P}_{1,-1,1,-1}\left(\omega\right)+\mathbf{P}_{1,-1,-1,1}\left(\omega\right)+\mathbf{P}_{1,1,0,0}\left(\omega\right)+\mathbf{P}_{0,0,1,1}\left(\omega\right) & =1, \label{eq:KS-omega-3}\\
\mathbf{P}_{1,-1,1,-1}\left(\omega\right)+\mathbf{P}_{1,1,1,1}\left(\omega\right)+\mathbf{P}_{1,0,-1,0}\left(\omega\right)+\mathbf{P}_{0,1,0,-1}\left(\omega\right) & =1, \label{eq:KS-omega-4}\\
\mathbf{P}_{0,0,1,0}\left(\omega\right)+\mathbf{P}_{0,1,0,0}\left(\omega\right)+\mathbf{P}_{1,0,0,1}\left(\omega\right)+\mathbf{P}_{1,0,0,-1}\left(\omega\right) & =1,\label{eq:KS-omega-5}\\
\mathbf{P}_{1,-1,-1,1}\left(\omega\right)+\mathbf{P}_{1,1,1,1}\left(\omega\right)+\mathbf{P}_{1,0,0,-1}\left(\omega\right)+\mathbf{P}_{0,1,-1,0}\left(\omega\right) & =1, \label{eq:KS-omega-6}\\
\mathbf{P}_{1,1,-1,1}\left(\omega\right)+\mathbf{P}_{1,1,1,-1}\left(\omega\right)+\mathbf{P}_{1,-1,0,0}\left(\omega\right)+\mathbf{P}_{0,0,1,1}\left(\omega\right) & =1, \label{eq:KS-omega-7}\\
\mathbf{P}_{1,1,-1,1}\left(\omega\right)+\mathbf{P}_{-1,1,1,1}\left(\omega\right)+\mathbf{P}_{1,0,1,0}\left(\omega\right)+\mathbf{P}_{0,1,0,-1}\left(\omega\right) & =1,\label{eq:KS-omega-8}\\
\mathbf{P}_{1,1,1,-1}\left(\omega\right)+\mathbf{P}_{-1,1,1,1}\left(\omega\right)+\mathbf{P}_{1,0,0,1}\left(\omega\right)+\mathbf{P}_{0,1,-1,0}\left(\omega\right) & =1.\label{eq:cabello-last-omega}
\end{align}
Now the contradiction  becomes clear: if we add all the random variables on the left hand side (which are $0$ or $1$ valued), we obtain an even number, since each $\mathbf{P}_{i}$ appears twice, whereas on the right hand side the sum adds to nine, which is an odd number.  However, notice that this contradiction only appears because we are assigning the same element $\omega$ of the sample space to, say,  $\mathbf{P}_{0,0,0,1}$ in (\ref{eq:cabello-1st-omega}) as well as $\mathbf{P}_{0,0,0,1}$ in (\ref{eq:KS-omega-2}). This is justifiable by an ontological assumption: if $\mathbf{P}_{0,0,0,1}$ is the property of a system, such property exists independent of what other properties we measure with it. However, as KS shows, this  assumption presents a challenge: in QM, properties are dependent of context.  

Contextuality is not a novel effect, as it is present in many experimental situations outside of quantum mechanics, but in quantum mechanics it takes a central role. However, we point out that, though the CbD approach is consistent and somewhat resolves the problems associated with properties in quantum mechanics, it is not clear to us what quantum ontology it suggests. Furthermore, the formalism of quantum mechanics, with properties $A$ being represented by the same Hilbert-space observable regardless of the experimental context, does not clearly distinguish between properties in one context or another. So, it is our goal here to provide an alternative approach for quantum properties that is consistent with the quantum formalism as well as free of inconsistencies with respect to a classical logic. We do so in Section \ref{sec:Quasi-Properties}, where we show how we could use quasi-sets to create random variables that can violate inequalities such as (\ref{eq:suppes-zanotti}). In Sections \ref{sec:KSH} and \ref{sec:qset_contradiction} below we return to the Kochen-Specker contradiction in order to see how it can be avoided using qsets. 

\subsection{Indistinguishability and contextuality}\label{sec:Quasi-Properties}\label{sec:iacmp}

Let us now interpret the KS theorem in a different way, using a notion of \textit{indistinguishability} (or \textit{indiscernibility}) for both particles and properties. The notion of indiscernibility is to be taken as intuitive for now, but it shall be made precise in the next sections. Here we aim mainly to motivate the most formal sections below. 

Intuitively, our interpretation  can be expressed by means of the following question: what if, instead of one identifiable particle or property, as we have considered above when we have taken $\omega \in \Omega$, we have an indistinguishable collection of them? Suppose we have a collection  of such indiscernible entities (as we shall see, we will express this by referring to such a collection as a \textit{quasi-set}, or just a \textit{qset} for short).  Without lost of generality or implying that we are supposing that we can speak of a \textit{difference} among then, by the lack of an adequate word, we shall refer to them as ``different.''\footnote{As Schrödinger have stressed regarding this subject, ``a particle is not an individual. (\ldots) it lacks sameness (\ldots) It is not at all easy to realize this lack of individuality and to find words for it.'' \cite{schroedinger_science_1952} In fact, we need to circumvent the difficulties with subterfuges of language, by using words like "identity" and "difference" which do not seem to conform with the situations.}  But we must insist that this \textit{façon de parler} can be made rigorous in terms of a quasi-set of indistinguishable objects representing quantum systems. So  we may reason as if we have a ``different'' particle or property for each context (that is, one \textit{counting as different}). 

This talk about ``difference'' is a way of speaking; they are indistinguishable but do not count as one. This agrees with the physicists' jargon, but not with the underlying mathematics (see below). As we have said, we have particles or properties in different contexts (seen as sets of properties), but we cannot say which particles or properties are in each context, since they are indistinguishable. The above examples illustrate the situation: assuming that properties are the same leads to contradictions. Thus, each particle is responsible for the outcomes in each contextual measurement. 

The fact that a "different" particle could be involved in a "different" context  (an indistinguishable one, for the considered properties are also indistinguishable from those of the first context) allows us to have different values for indistinguishable particles. This is the fundamental point: we have indistinguishable properties (to be defined in the next section) and indistinguishable particles. Take a collection (qset) of such properties: this is a context. We may form several contexts this way. Take a particle and one context and measure the corresponding properties: we have outcomes. Now take another "indistinguishable" context and an indistinguishable particle. Although the properties \textit{and} the particles are indistinguishable, the outcomes may be different.

\section{The ontological thesis and its mathematics}\label{sec:ontology}

We emphasized before some peculiarities of quantum systems which may call our attention to a deeper look to phenomena such as KS. In this section we attempt to justify our  thesis about the ontology of these quantum entities, where ``entity'' is used here as synonymous of ``thing'', ``object'' and other terms which refer to the entities we are interested in.  Let us first fix some terminology, to be further explained in Section \ref{identitydifference}. By an \textit{individual} we informally mean an entity that \textit{possesses identity}, in the sense of being able to be identified as such in a certain circumstance and as \textit{the same entity} we have had enrolled with in another circumstance (e.g.\  we recognize Mr. Bean when he appears on TV in every instance of his appearance).\footnote{David Hume, in his \textit{Treatise on Human Nature}, has cast doubts even on this view, for according to him there is nothing except pure habit that assures us that the Mr. Bean of today is \textit{the same} Mr. Bean of some days before.} Of course this is an ontological thesis, one we tend do accept without further discussion: an individual is identical just to itself and to nothing more. This identity is called \textit{self-identity}, or \textit{numerical identity}, to distinguish it from the \textit{relative identity} we shall mention below. 

It is important to mention that, in assuming the non-individuals view, we are not positing that there are no entities at all, or that all things are merged in a great fuzzy swarm of something. As we shall realize soon, these non-individuals can be put apart or \textit{isolated}. The terminology `non-individuals' might not be good, but as we have seen it has a long tradition. It would be better perhaps to call them, as did Weyl, \textit{individuals without identity} (an oxymoron), or something like that, but we shall continue to follow the tradition and refer to them as non-individuals.

In a formal setting, we can say that an individual is something that obeys the laws of the classical theory of identity embedded in classical logic (either of first or of higher-order, that is, set theory --- for details, see \cite[Chap. 6]{french_identity_2006}). An individual is \textit{different} from every ``other'' individual: there cannot be two individuals completely similar, without a difference. If they are two, they must present an internal (not spatial) difference, something that today we would think of  in terms of intrinsic properties (more on this below). This is the famous Leibniz's Principle of the Identity of Indiscernibles, part of Leibniz's metaphysics which was incorporated in classical logic, standard mathematics, and classical mechanics. If we use standard mathematics---say one that can be constructed within a standard framework like the Zermelo-Fraenkel set theory---then all entities are individuals:  given an  $a$, define the ``property'' $I_a$ as \textit{be identical with $a$} as follows: $I_a(x) \leftrightarrow x \in \{a\}$ (considering that the unitary set does exist for any $a$). Then, $I_a$ is a property shared only by $a$, and, by Leibniz's principle, any other object will be different from $a$ for not having this property. This is the core of what we can call \textit{classical metaphysics}: it is a metaphysics of individuals, as in classical mechanics, where by hypothesis any particle can be discerned from any other (even of the similar species) by their trajectories at least.\footnote{It is interesting to note that, although being of the same species and thus partaking all their intrinsic properties, two classical particles are regarded as distinct. Some, like Heinz Post, say that they present a `transcendental individuality' beyond their attributes, but this is to push metaphysics too far -- see \cite{french_identity_2006}.}

Ontologically speaking, the formalism of non-relativistic quantum mechanics (QM) is compatible with more than one view (see \cite{french_identity_2006} for an extensive discussion). This is termed the \textit{underdetermination of the metaphysics by the physics} (ibd., \S 4.5). There are two main ontological views that have been developed in the literature. The first is the \textit{Received View} (ibid., p.135), for it has its origins with the forerunners of QM, specially Schrödinger, Heisenberg, and Weyl. This is the view which starts with the idea that, in the quantum realm, particles (and, of course, other quantum systems) \textit{lose their individuality}, since in most situations we cannot identify them as individuals anymore. For example, Schrödinger said that  
\begin{quote}
``[We are] compelled to dismiss the idea that (\ldots) a particle is an individual entity which retains its `sameness' forever. Quite on the contrary, we are now obliged to assert that the ultimate constituents of matter have no `sameness' at all. 

(\ldots)

I beg to emphasize this and I beg you to believe it: It is not a question of our being able to ascertain the identity in some instances and not being able to do so in others. It is beyond of doubt that the question `sameness'’ of identity, really and truly has no meaning.'' \cite[pp.117-8]{schroedinger_science_1952}, \cite[p.119]{french_identity_2006}
\end{quote}
Weyl (and Heisenberg) goes in the same direction, writing, for instance, that ``photons (\ldots) are individuals without identity"\ \cite[p.246]{weyl_philosophy_1949}, \cite[\S 3.7]{french_identity_2006}, using a confusing terminology, since individuals are entities which \textit{do have} identity. Unsurprisingly, this view has also been called the \textit{non-individuals} view. 

The second view considers quantum particles as individuals, similar to classical particles, thus associating QM to a ``classical'' ontology, taking quantum entities as individuals on pair with their ``classical twins.'' In this view, we need to impose particular restrictions to the states particles may be in: either symmetric or anti-symmetric states for ordinary particles, but not other states formed by particle permutations corresponding to paraparticles \cite[\S 4.1.2]{french_identity_2006}. For example, Bohm's interpretation of QM starts from the supposition (a metaphysical one) that particles are individuals as well. 

Here we pursue the non-individuals view. Our reasons can be stated in two different ways. The first is that most interpretations of QM seem to favor this view instead of the view of quantum entities as individuals. Second, as we shall summarize soon, there are situations where it makes no sense to claim that the involved systems can have the characteristics of individuals as posed above. For instance, there is no way of discerning between two entangled particles, or among the particles/atoms that form a Bose-Einstein condensate.

The important analogy for our purposes may be this. Take a chemical reaction, such as methane combustion: 
\begin{equation}
CH_4 + 2 O_2 \to CO_2 + 2 H_2O.
\end{equation}
There are four Hydrogen atoms in the methane molecule, and no differentiation among them is possible. After the reaction they are part of the two water molecules. But, which ones are where? It is impossible to say. The same can be said about the four Oxygen atoms. If we were able to discern then, we would be introducing some additional property (a quantum number) of the elements that chemistry says they do not possess. It is in this sense that we speak of non-individuals: we have ``entities,'' quantum entities for the lack of a better expression, but according to us they should be taken as devoid of identity. It is this view we wish to explore. 

But physics, and its ontology, need to be described mathematically. How can we describe our objects without identity conditions? Of course we cannot use standard mathematics, for the objects it describes, as mentioned above, are \textit{individuals}, and the only way to consider non-individuals in this framework would be by confining them within a non-rigid structure, that is, a mathematical structure having automorphisms other then the identity function. Let $\mathfrak{A} = \langle A, R_i \rangle$ be a structure, being $A$ a non-empty set and  $R_i$ an $n$-ary relation over the elements of $A$, where $i$ range over a set of indexes. An automorphism of $\mathfrak{A}$ is a bijection $h : A \to A$ such that $R_k(x_1,\ldots,x_n) \leftrightarrow R_k(h(x_1),\ldots,h(x_n))$ for every $R_k$. Of course the identity function is an automorphism, and if it is the only one, the structure is called \textit{rigid} or \textit{not-deformable}. Two elements $a, b \in A$ are $\mathfrak{A}$-indiscernible if there is an automorphism $h$ such that $h(a)=b$, otherwise, they are $\mathfrak{A}$-discernible. For instance, $i$ and $-i$ are indiscernible in the structure of the field of complex numbers, for the application which associates a complex number to its conjugate is an automorphism. 

So, in considering non-rigid structures, we can talk about indiscernible objects and perhaps even about non-individuals. But there is a fallacy here in the case of non-individuals: the indiscernible objects are in fact \textit{fake non-individuals}, for they act as such only inside of the structure. \textit{From the outside}, that is, from the point of view of the whole set theory, which by hypothesis we assume is the mathematical basis of our physical (or ontological) theory, we can see that they are individuals. But we always can go \textit{outside} of a structure, as da Costa and Rodrigues \cite{costa_definability_2007} proved in a theorem: in the scope of standard set theories like  Zermelo-Fraenkel, \textit{every structure can be extended to a rigid one}. This means that even if from inside a structure we cannot distinguish the elements, from the outside we always can. For example, we know that $i \not= -i$, but we cannot realize that from inside the structure of the complex numbers. This is why we say that the entities described by classical mathematics are individuals, and this is the reason why we need to pursue a different, non-classical, mathematical setting, if our aim is to deal with \textit{legitimate} non-individuals. This is what we consider next.

\subsection{Indistinguishability and quasi-sets}\label{sec:qsets}

To motivate the mathematical framework, we start by  describing informally the notion of indistinguishable properties we use in this paper. Later on, this notion will be formally described in the \textit{theory of quasi-sets} (or \textit{qset theory} for short).  We shall be  speaking informally of ``indistinguishable'' (or ``indiscernible'') things, as well as of ``identical''  and ``different'' things. These concepts must be taken, at this stage, in their intuitive sense---the theory will make them precise. Our first discussion of those basic concepts may seem nonchalant, but we ask you, the reader, to be forgiving with us at this stage, for we need to bring your attention for some details which in the standard setting are assumed as quite obvious, but which need to be taken if we are to have truly indistinguishable things.\footnote{There are two main lines of assumptions which define the theories conferring identity to an object: the \textit{substratum} theories, which postulate the existence of something beyond the properties of an entity, and \textit{bundle theories},  which say that the characteristics of the objects, including its identity, are done by some of its attributes. The discussion involving these views, mainly related to quantum mechanics, can be seen in \cite{french_identity_2006}.} In the subsequent discussions we need the following the definitions:
\begin{description}
	\item[Indistinguishable things.] \textit{Indistinguishable} things are things that share all their properties. For example, two photons prepared in the same state in a cavity cannot in any way be distinguished. 
	\item[Relative indistinguishable things.] \textit{Relative} indistinguishable things are things which partake some attributes, those relative to which they are indistinguishable.  Classical Newtonian particles, for instance, may be indistinguishable  with respect to all their attributes  (e.g., mass, charge, etc), but are distinguishable by their trajectories. 
	\item[Identical things.] \textit{Identical} things are the very same thing.  Again, from classical physics, a particle at time $t_1$ is identical to the same particle at time $t_2$. 
    \item[Different things.] \textit{Different} things are things which present a difference, an attribute of its own which confers it an \textit{identity} not shared by any other entity.\footnote{We are avoiding any compromise with substance, \textit{haecceities} and so on \cite{french_identity_2006}} An example would be two classical particles with different masses, i.e. one with mass $m$ and another with mass $M\neq m$. 
 \end{description}
 
\subsubsection{``Another'' electron and the motivation for the theory \texorpdfstring{$\mathfrak{Q}$}{Q}}\label{identitydifference}

Let us consider the case of a Helium atom in its fundamental state. This atom has two electrons, and if we measure their spins in a certain direction $z$, we will find one ``up'' and another ``down.'' But nothing allows us to distinguish between the electrons: we cannot say which is which, and in fact this question may even be meaningless. In the quasi-set theory $\mathfrak{Q}$, we may speak of, say, a qset $X$ as having q-cardinal equal to 2, but still having its elements  indistinguishable. Talking about $X$ replaces talking about anti-symmetric functions (in the case of electrons), but the conclusions are the same: there is nothing in $\mathfrak{Q}$ that enables us to distinguish between two elements of $X$. So, this partially captures what the standard view of indistinguishability in QM says.

As we have emphasized, we need to speak of the \textit{another} electron in the He atom. This is, again, a \textit{façon de parler}. In order to make this precise, we need to consider the notion of \textit{difference}, that is, the very notion (theory) of \textit{identity} needs to be considered (as difference is the negation of identity). The informal idea of identity that interests us here  is that of \textit{numerical identity}: a thing has (numerical) identity if counts as one, that is, if it has an identity card, something which even if only in principle enables us to discern it from any other thing in whatever situation, even if it is mixed with other similar things of the same kind. We say that such an entity is an \textit{individual}. Sometimes we cannot make the difference explicit due to lack of experimental accuracy or other empirical difficulties, but the possibility exists in principle. This is what happens with \textit{classical} particles and with all entities considered by standard mathematics. 

In classical physics, two particles of the same kind may be indistinguishable relative to all properties (\textit{relative identity}, not numerical identity), but their positions in space and time discern them one from each other in any situation. All impossibility in making the distinction must be regarded as an \textit{epistemological} ignorance, but they do present identity, they are ontologically distinct things. Identity makes sense in this classical realm.

Classical logic, standard mathematics, and classical physics were built with this informal notion of identity in mind, which we suppose (this is of course a metaphysical thesis) also applies to  objects in our surroundings. Identity as such is well summarized by Leibniz's Principle of the Identity of Indiscernibles: if we have \textit{two} things, each one presents a property not shared by the another. Conversely, if they present all their properties as common, that is, if they are \textit{indistinguishable}, then they are the very same object, in the sense that there is not more than one object, just one. The use of these terms in physics and in philosophy vary. \textit{Identity} in quantum mechanics means agreement in all the intrinsic properties \cite[p.275]{jauch_foundations_1968}. But, in philosophy,  so as in standard mathematics, identity means \textit{the same}. If $a=b$ is true, this means that there are not two distinct things, but just one, which can be named either by $a$ or $b$. Physicists understand the distinction between one and the other, but their language should be made more precise in mathematical and logical terms. That's what we partially intend to do here.

Classical physics, as we have seen, classical logic, and, more importantly, standard mathematics do not enable indistinguishable but not identical things, as the formalisms involve Leibniz's metaphysics of identity in some way (the converse of the above principle is a theorem of the underlying logic: identical things are indistinguishable; they have the same properties). So, how can we speak of indistinguishable but not identical things? One way is to relax the notion of identity and consider just \textit{relative identity}, that is, identity relative to a certain class of properties. For instance, the three authors of this paper are indistinguishable relatively to their interest for physics, mathematics, philosophy, and good beer, but they are not identical, as they are not the same person. In being different, each of them present at least property (in our case, a large quantity of them) which do not belong to the others, e.g. the countries we live in, US, Argentina, and Brazil. So, within a ``classical'' setting such as classical mathematics\footnote{What is ``classical'' mathematics is unclear, but here we mean a mathematics that can be constructed within a standard set theory, such as the Zermelo-Fraenkel system.}, we can deal with indistinguishable things only by pretending that they share \textit{all} their properties  (and say that they are "indistinguishable"), but this is  (again!) a way of talking---in such a framework, every object is an individual, as it has identity.\footnote{In ZF, given an object $a$, it is enough to consider its unitary set $\{a\}$ and define the property  $I_a(x) := x \in \{a\}$. It is a theorem of ZF that the only object having the property $I_a$ is $a$ itself. Then, according to Leibniz's principle, this suffices for saying that $a$ is \textit{distinct} from \textit{any} other object.}

Quasi-set theory is the mathematical theory of indistinguishable but not identical things. Why to consider it? The reason comes, first and foremost, from QM: to express  without making the trick of using a relative identity (say by making use of an equivalence relation or a congruence other than identity), such that we may have ``truly'' indistinguishable things (quantum systems) that cannot be discerned \textit{even in principle}. Cases abound in quantum mechanics: the aforementioned two electrons in a  He atom, and particles in a state of superposition, to name a few. In quantum field theory, the Bose-Einstein condensate is perhaps the best example of ``things'' (quantum systems) partaking all their properties, being in the same quantum state, without turning to be, by force of Leibniz's principle, the very same object. Notice that this is not an epistemic indistinguishability: quantum particles are indistinguishable in principle. So, to make things mathematically right, we need of such a theory, and we need to change the metaphysics accordingly.\footnote{It is a thesis of ours that we cannot read the metaphysics from the physics. There is an \textit{underdetermination} of the metaphysics by the physics. See reference \cite[\S 4.5]{french_identity_2006} for details.} In QM, Leibniz's principle would not be valid for all objects, yet we can maintain it for some others. In $\mathfrak{Q}$, which we explain in the next section without to many technical details (for the interested reader, see \cite{french_identity_2006,french_remarks_2010}), the standard theory of identity does not hold for all objects of the domain.

\subsection{Quasi-set theory}\label{qsettheory}
Now that we laid out the motivations, let us have a look on how the qset theory describes the above concepts. Indiscernibility is a primitive concept, formalized by a binary relation ``$\equiv$'' satisfying the properties of an equivalence relation. In this notation, ``$x\equiv x^\prime $'' is thought to mean "$x$ is indiscernible from $x^\prime$''. This binary relation is a partial congruence in the following sense: for most relations, if $R(x,y)$ and $x \equiv x^\prime$, then $R(x^\prime, y)$ as well (the same holds for the second variable). The only relation to which this result does not hold is membership: $x \in y$ and $x^\prime \equiv x$ does not entail that $x^\prime \in y$ (details in \cite{french_identity_2006,french_remarks_2010}). This captures the idea that, although two electrons are indistinguishable, one of them may be in an orbital, while the other may be not. We should emphasize that it is wrong to conclude that since they are in different orbitals, they are distinct, for one has a property not shared by the other, namely, to belong to the orbital. In fact, we cannot know which is which and, furthermore, any permutation of the electrons does not give different empirical results. 

This conclusion that being in different orbitals would lead to electrons being different holds only if the metaphysics is classic, that is, only if we reason \textit{as if} the systems are classical objects having identity, that is, obeying the standard theory of identity. But our theory says that for $q$-objects the predicate of identity does not hold. For instance, it makes no sense to say that the two electrons in the level 2$s$ of a Sodium atom are distinct (with orbitals $1s^2 2s^2 2p^6 3s^1$), for they lack a distinctive property that enables us to apply Leibniz's principle and conclude that they are different. The reason is that even if they differ in their values of spin (being fermions, they cannot be in the same quantum state), it would make sense to talk about their identity only if we could identify them. For example, we can say that one has spin UP, but we cannot say which one. Hence, they cannot present identity conditions.\footnote{The situation is different with ``classical'' objects. For instance, in Zermelo-Fraenkel with the Axiom of Choice, every non-empty subset of the real numbers admits a well-ordering. So, the last elements of two disjoint subsets are \textit{different}, although we cannot say which reals they are, for the well-ordering cannot be defined by any formula of the language of set theory. But, as they obey the classical theory of identity, they are \textit{in principle} different, yet unidentifiable.}

The objects of the considered domain are distinguished as $q$-objects (intended to represent quantum objects), $c$-objects (representing classical objects), and collections of them, termed \textit{quasi-sets} (qsets), some of them perhaps being mixed collections, yet these are not the interesting ones. Among the qsets there are some called \textit{sets}, which have as elements either $c$-objects or other sets. The null qset is a set. The $q$- and $c$-objects are ur-elements, in the sense of the set theories with objects which are not sets but which can be elements of sets \cite{suppes_axiomatic_1972}. If we eliminate the $q$-objects, we are left with a copy of ZFU, the Zermelo-Fraenkel set theory with \textit{Urelemente}. Hence, we can reconstruct all standard mathematics within $\mathfrak{Q}$ in such a ``classical part'' of the theory. 

Cardinals are also taken as primitive, although they can be proven to exist for finite qsets (finite in the standard sense \cite{domenech_discussion_2007}).  The idea is to use this concept to enable us to speak of ``several objects'' in a certain situation and expressing that in terms of cardinals. So, when we say that we have two indiscernible q-functions, according to the above definition, we are saying that we have a qset whose elements are indiscernible q-functions and whose q-cardinal is two\footnote{We use the notation $qc(x)=n$ (really, $qc(x) =_E n$, see below) for a quasi-set $x$ whose q-cardinal is $n$.}. The same happens in other situations. 

The interesting fact is that qsets composed by several indistinguishable objects do not have an associated ordinal. This means that these elements cannot be ordered, hence they cannot be counted. But even so we can still speak of the cardinal of a collection, termed its \textit{quasi-cardinal} or just its \textit{q-cardinal}. This is similar to what we have in QM when we say that we have some quantity of systems of the same kind but cannot individuate or count them, e.g. the six electrons in the level 2$p$ of a Sodium atom (cf. above). 

Identity (termed \textit{extensional identity}) ``$=_E$'' is defined for qsets having the same elements (in the sense that if an element belongs to one of them, them it belongs to the another)\footnote{There are subtleties that require us to provide further explanations. In $\mathfrak{Q}$, you cannot do the math and decide either a certain $q$-object belongs or not to a qset, for this requires identity---you need to identify the object you are making reference to. In the theory, however, you can make the hypothesis that \textit{if} a certain object belongs to a qset, then so and so. This is similar to Russell's use of the axioms of infinite ($I$)  and choice ($C$) in his theory of types, which assume the existence of certain classes that cannot be constructed, so going against Russell's  constructibility thesis. 
What was Russell's answer ? He transformed all sentences $\alpha$ whose proofs depend on these axioms in conditionals of the form $I \to \alpha$ and $C \to \alpha$. Hence, \textit{if} the axioms hold, \textit{then} we can get $\alpha$. We are applying the same reasoning here: \textit{if} the objects of a qset belong to the another and vice-versa, \textit{then} they are extensionally identical. It should be noted that the definition of extensional identity  holds only for sets and for $c$-objects.} 
or for $c$-objects belonging to the same qsets. It can be proven that this identity has all the properties of classical identity for the objects to which it applies. But it does not make sense for $q$-objects, that is, $x =_E y$ does not have any meaning in the theory if $x$ and $y$ are $q$-objects. It is similar so speak of categories in the Zermelo-Fraenkel set theory (supposed consistent). The concept cannot be captured by the theory, yet it can be expressed in its language. From now on, we shall abbreviate $=_E$ by $=$ as usual.

The postulates of $\mathfrak{Q}$ are similar to those of ZFU, but by considering that now we may have $q$-objects. The notion of indistinguishability is extended to qsets by means of an axiom which says that two qsets with the same q-cardinal and having the same quantity (we use q-cardinals to express this) of elements of the same kind (indistinguishable among them) are indiscernible too. As an example, consider the following: two sulfuric acid molecules H$_2$SO$_4$ are seen as indistinguishable qsets, for both contain q-cardinal equals to 7 (counting the atoms as basic elements), and the elements of the sub-collections of elements of the same kind are also of the same q-cardinal (2, 1, and 4 respectively). Then we can say that H$_2$SO$_4$ $\equiv$ H$_2$SO$_4$, but of course we cannot say that H$_2$SO$_4$ $=$ H$_2$SO$_4$, as for the latter the two molecules would not be two at all, but just the same molecule. In the first case, notwithstanding, they count as two, yet   we cannot say which is which.

Since we want to talk about random variables over qsets, it is important to define functions between qsets. This can be done in a straightforward way, \label{qfunctions} and here we consider binary relations and unary functions only. Such definitions can easily be extended to more complicated multi-valued functions. A (binary) q-relation between the qsets $A$ and $B$ is a qset of pairs of elements (sub-collections with q-cardinal equals 2), one in $A$, the other in $B$.\footnote{We are avoiding the long and boring definitions, as for instance the definition of ordered pairs, which presuppose lots of preliminary concepts, just to keep with the basic ideas. For details, the interested reader can see the indicated references.} Quasi-functions (q-functions) from $A$ to $B$ are binary relations between $A$ and $B$ such that if the pairs (qsets) with $a$ and $b$ and with $a^\prime$ and $b^\prime$ belong to it and if $a \equiv a^\prime$, then $b \equiv b^\prime$ (with $a$'s belonging to $A$ and the $b$'s to $B$). That is, a q-function may take indistinguishable elements to indistinguishable elements. When there are no $q$-objects involved, the indistinguishability relation collapses in identity and the definition is equivalent to the classical one. In particular, a q-function from a `classical' set such as $\{1,-1\}$ to a qset of indiscernible q-objects with q-cardinal $2$ can be defined so that we can't know which q-object is associated to each number (this example will be used below). 

To summarize, in this section we showed that the concept of indistinguishability, which is in conflict with Leibnitz's Principle of the Identity of Indiscernibles, can be incorporated as a metaphysical principle in a modified set theory with indistinguishable elements. This theory contains in it `copies' of the Zermelo-Frankael axioms with \textit{Urelemente} as a special case, when no indistinguishable $q$-objects are involved. This theory will provide us the mathematical basis for formally talking about indistinguishable properties, which we will show can be used in a theory of quantum properties.  We will see in the next section how we can use those indistinguishable properties to avoid contradictions in quantum contextual settings such as KS. 

\section{The indistinguishability assumption, contexts, and the measurement process}\label{sec:indisting-assumption}

Let us now relate the above metaphysical discussion to physics. Suppose that we aim to perform a quantum experiment. In order to check some statistical predictions of the formalism, we need to repeat the same experiment a number $N$ of times, with $N$ large enough to allow us to compute mean values, probabilities, and all necessary stochastic properties of experimental outcomes. But what can we mean by ``the same experiment?'' Let us elaborate on this notion to underscore how indistinguishability is deeply connected to contextuality.
 
First of all, notice that in order to make $N$ experiments, we must first prepare $N$ "identical" copies of a quantum system. That is, by employing the language of  most physicists, we need $N$ "identical" particles or quantum systems. But, according to the indistinguishability postulate,\footnote{This assumption, essential in the standard quantum formalism, says that the expected value of the measurement of any observable in a system in a given state is the same before and after the system has suffered an interchange of indistinguishable particles. Formally, $\langle \psi\ | \hat{A} | \psi\rangle = \langle P\psi | \hat{A} | P\psi\rangle$ for any observable $\hat{A}$ and any state $|\psi\rangle$, where $P$ is a permutation operator---see \cite[p.135 and \textit{passim}]{french_identity_2006}.} these particles cannot be identified. As it is generally agreed, this is so even if we perform a thought experiment: if we want to respect what the logic of QM seems to suggests (at least to us) with regards to identity, we must assume that the set of copies that we imagine of the quantum system is in reality a quasi-set of indiscernible objects in the sense of Section \ref{sec:qsets} above.
 
Next, we have to perform the ``same'' measurement (or more generally, the same set of measurements) on each preparation. This means to construct the "same" experimental setup for each one of them, which is impossible to realize. But  we can suppose that this construction  involves equivalent setups $M_i$, ($i=1,\ldots,N$), which are essentially \textit{indistinguishable} between them.  Notice that each one of these setups defines the ``same'' context. The fact that these setups are macroscopic should not lead us into confusion about their indistinguishable logical nature. This assumption  is inherited from the fact that particles are indistinguishable: as representative of properties, the $M_i$'s are indistinguishable in the sense given in Section \ref{sec:qsets}.
 
But our above discussion has a direct connection to the KS contradiction: when we run each version of the experiment, we may obtain different outcomes, even if we measure then in the same context. For example, if we prepare $N$ copies of a spin $1/2$ system and measure the spin in the same direction, say $S_z$, we can obtain a distinguishable series of results. As an example with $N=5$, we may obtain $(1/2,1/2,-1/2,-1/2,1/2)$. But here it comes the interesting part: while all preparations and measurements are essentially equivalent (i.e., indistinguishable), they are not the \textit{same} ones in the sense of being just one.  This is what allows a quantum system to possesses different results for equivalent experiments and still maintain an ontology based on truly indistinguishable entities. 
 
It is interesting to consider the classical analogue of this problem. If we prepare $N$ classical particles in the same state, from a logical perspective, the ontological properties of classical identity imply that, if we perform the same measurements on each particle, we must obtain the same results. There is no other logical possibility: \textit{two classical particles prepared in the same state (even if there are actually two of them), are identical from the logical perspective---classical logic of course}. Thus, if we prepare identical (equivalent) measurements we must obtain (ideally) the same results. Under these assumptions, statistical fluctuations need to be seem as originated by the imperfections of either the state preparation or measurement, but never in the ontological properties of the objects themselves. They are, at least in principle, well behaved \textit{individuals}. That is, there is no room for fluctuations at the logical/ontological ``classical'' level: the classical theory of identity implies that indistinguishability must collapse into identity (in the philosophical or in the mathematical sense of ``being the same'').
 
The quantum mechanical situation is totally different from a classical logical/ontological point of view, provided that we assume that quantum particles can be truly indistinguishable objects. By this we mean: contrarily to classical systems, quantum systems can be different {\em solo numero}, i.e., they can be seen as  collections of indistinguishable entities in a very strong sense which, notwithstanding, are not the same. This logical structure does not allow us to conclude that, in a thought experiment, the results of the $N$ measurements $M_i$ must be the same. With this in mind, there is  room for another possibility: due to the fact that particles can be seen as truly indistinguishable entities, they are not obliged to yield exactly the same results even if the experiments are indistinguishable. Suppose that we aim to perform a quantum experiment. A quantum property, in this sense, cannot truly be attributed to a particle, since this particle does not have an identity. 

This idea that measurements may yield different results for different but indistinguishable particles may suggest the possibility of distinguishing them, since we could label them with the measured property. This is not so, as we cannot attribute each result of the experiment to each particle (before measurement), because of the fact that the state has changed, and the particles could have been even destroyed during the measurement process. In other words: the result of the measurement must not be confused with the particles themselves. And this is expressed also in the possibility of correlations between ``different'' particles. For example, two entangled spin-$1/2$ particles in a singlet state will show strong spin-measurement correlations, and if we measure their spin in the same direction they will be anti-correlated. However, even in this case, we cannot say that particle $a$ has spin ``up'' and $b$  has spin ``down'', as this is not possible within the theory.  All we can say, in this case, is that one of the particles has spin  ``up'' and  the other ``down''.

\subsection{The indistinguishability assumption and KS}\label{sec:KSH}

Let us now take indistinguishability, as presented above, as a metaphysical thesis and consider its implications for quantum contextuality. In particular, we examine the implications of indistinguishable objects to the KS contradiction, something we already suggested intuitively in the beginning of this section. As seen in Section \ref{sec:KS-argument}, we can summarize the assumption leading to the KS contradiction as the following statement:
\begin{quote}
(KSH) \;  It is possible to assign well-definite values to all measurable properties of a given quantum system.
\end{quote}
We can avoid a contradiction by negating KSH in at least two ways:  
\begin{description}
\item[(i)] properties do not have well-defined values 
\item[(ii)] properties or particles may be indistinguishable. 
\end{description}
For (i), given a  quantum system, it is \emph{not} possible to assign well-definite values to all  measurable properties. This is the usual way to avoid  the KS contradiction, as discussed in Section \ref{sec:KS-argument}.  Option  (i) is the most popular interpretation of the KS contradiction among physicists and philosophers of physics. Option (ii), as we shall see below, is a consequence of the indistinguishability of quantum particles, and has also been explored by Kurzynski \cite{kurzynski_contextuality_2017}. If particles are truly indistinguishable entities, and more than one particle could be involved in a measurement process of a quantum system, then the intrinsic lack of identity of particles makes it meaningless to speak about properties as being properties of a specific particle. 

In this paper we take option (ii), and analyze its possible consequences. This allows us to introduce a novel interpretation of the KS result.  In order to proceed, let us assume that quantum systems (or even properties) lack identity (in the sense explained above): they may be taken, in certain situations, as truly indistinguishable objects, and in these cases it is meaningless to label them, to name them, or to identify them. Notice that the entities involved need not  be particles: they can be degrees of freedom or even be fields. Only an \textit{indistinguishable ``thing''} is needed for our argument.

Under this non-individuality assumption, it seems odd to affirm that the properties defining a context $C$ correspond to the same particle than the ones defining a "different" context $D$ prepared in the same way (an indistinguishable context, in the sense posed before). The act of choosing between measurements in contexts $C$ or $D$ corresponds to different (and usually, incompatible) possible worlds, and we cannot grant that we are talking about one and the object underlying these alternatives: our non-individuality assumption implies that it is meaningless to assign transworld identity to elementary particles. Notice that this argument needs not to be operational: it follows as a logical consequence of our ontological non-individuality assumption. There is no need to perform any actual experiments in order to realize that to affirm that we have the same particle in all contexts is a strong ontological assumption (dependent on the classical notion of identity). 

In order to claim that a classical system possesses context-independent properties, we must be able to identify the system in different possible worlds first (from a logical/ontological point of view). Indeed, we could (trivially) simulate a false contextuality experiment using ``different'' (remember the restrictions posed above on the use of language) classical particles  (in the sense that we have a quasi-set with q-cardinal greater than one). But our fraud could always---at least in principle---be debunked. In classical mechanics, if we take our particle and measure a collection of properties (a context), and then measure another context, we can in principle follow the trajectory of our particle and assure that it will be \textit{the same} in the new context. This is why we cannot reproduce quantum contextuality in classical mechanics without forbidding the possibility of detecting the fraud: if someone tries to reproduce a contextuality experiment using different classical particles, it could always be debunked by a careful observer following the particles’ trajectories and denouncing that he have used more than one particle. But as Schrödinger observed (and following our non-individuality assumption), it is pointless to try to identify particles in the quantum setting: besides the fact that particles are usually destroyed or perturbed in a quantum experiment,  we have no means to detect which is which and “debunk” the false experiment  \cite[\S 3.6]{french_identity_2006}. And, we emphasize it again, this is not a matter to be settled in an empirical way: these considerations follow as a consequence of our ontological non-individuality assumption. If our particle is a quantum system, there is no way, when we repeat the experiment with another set of properties, to grant that we will be dealing with  \textit{the same} particle or property, just because elementary particles are (in certain crucial situations) indistinguishable and the very question is meaningless (according to our ontology, there are no identity conditions to elementary particles and their collections).

We remark that the above considerations do not imply that quantum particles are distinguishable by their (differing) properties in different contexts. Let us consider the example of a singlet state to illustrate this. In a singlet state, we know that electrons have opposite spin values: if we measure the spin in a certain direction, we obtain that one electron has spin up and the other will have spin down. And there is no other possibility (because of the properties of the singlet state). But this does not allow us to conclude a statement such as “electron 1 has spin up while electron 2 has spin up”. We can only conclude that electrons have different spins, and that is all. This example shows that

\begin{quote}
We can have  indistinguishable particles that although being of the same kind, form a quasi-set with q-cardinal greater than one and its elements may have different properties, while at the same time,
these particles do not have properties which allow us to individuate o identify them.
\end{quote}

There are many possible examples of this situation. Take again the six electrons in the 2$p$ level of an Sodium atom 1$s^2$2$s^2$2$p^6$3$s^1$. They are indiscernible and, although obeying Pauli's Principle in not having all the same quantum numbers, nothing can discern them, tells us which is which. Even so they have "different" properties, characterized for instance by their quantum numbers. The six electrons cannot be \textit{counted} if by this we understand, as in standard mathematics, to define a bijection from  the von Neumann ordinal number $6 = \{0,1,2,3,4,5\}$ into that collection. To which electron should we associate the number 4? Impossible to say. The most we can say is that we associate $4$ to \textit{one} of them, but without identification, something that can be captured by the use of q-functions. So, a \textit{standard} function cannot be defined, and this is why we use q-functions. 

\subsubsection{Quasiset theory used to avoid the contradiction}\label{sec:qset_contradiction}

To begin, let us mention something more about the defined notion of \textit{extensional identity} given in  $\mathfrak{Q}$.  It says that two items $x$ and $y$ are extensionally  identical ($x =_{E} y$, abbreviated b $x = y$) if  they are both $q$-objects and belong to the same qsets or are qsets having the same elements (the formal definition is given  in  \cite[p.277]{french_identity_2006}). If there are no $q$-objects involved, the definition collapses in the standard definition of identity in ZFC. 

Now, from the arguments exposed above, it follows that if we assume that particles can be truly indistinguishable entities, the contradiction in the KS theorem can  be avoided, so it seems. Let us now use quasi-set theory to express this idea in a formal way. 

The first mathematical notion that we need is that of a \textit{strong singleton}. Given a qset $z$ and a $q$-object $x \in z$, we can always form (by the qset version of the separation axiom) the  qset $[x]_z$ of all elements of $z$ that are indistinguishable from $x$ (we follow the notations introduced in \cite{french_remarks_2010}).  Then, again by separation, we get the  \textit{strong singleton} of $x$, written 
 $\llbracket x \rrbracket_z$ as a subqset of $[x]_z$ having q-cardinal equals to one.  That is,  $\llbracket x \rrbracket_z$ is a qset whose q-cardinal  is one and whose only element is an indistinguishable from $x$. We cannot say that this element \textit{is} $x$ for  this affirmation presupposes identity. The existence of such a qset results from the axioms of the theory, as it was shown  in  \cite[p.292]{french_identity_2006}. 

So, $\llbracket x \rrbracket_z$ represents a class of objects, rather than a single object, and satisfies the following property:
\begin{equation}
(\llbracket x \rrbracket_z \subseteq[x]_z)\wedge\mbox{qc}(\llbracket x \rrbracket)=_{E}1
\end{equation}
Notice once more that, despite the notation, it is impossible to identify which is the element that belongs to $\llbracket x \rrbracket_z$. Any indistinguishable from $x$ will do the job and the question is simply meaningless inside quasi-set theory, due to the fact that the standard theory of identity does not apply to $q$-objects. Furthermore, we remark that all the elements (strong singletons) of such a class are also indistinguishable. 

Now, let us consider each projection operator involved in the above equations expressing KS. Each projection is of the form $\hat{P}_{i,j,k,l}$, where $i,j,k,l$ take values in the set $\{0,\pm 1\}$, and to such projectors we associate a property $P_{i,j,k,l}$. Each collection of values $(i,j,k,l)$ represents a possible empirical proposition. Now, on each run of the experiment, these propositions can be either true or false. This is expressed formally by assigning to a random variable $\mathbf{P}_{i,j,k,l}$ the value $1$ if the proposition is true and the value $0$ if it is false. Here we stress that such a proposition refers to an identifiable particle. 

In the case when the particle is treated as a ``classical particle'',  let us call it $e$ (in this case, $e$ can be taken as an element of the \textit{classical part} of quasi-set theory, so it is governed by the classical theory of identity). For example, if we check the property $P_{1,0,0,0}$ and it is true, we describe this by the proposition: ``the particle $e$ possesses the property $P_{1,0,0,0}$". This situation can be represented by the ordered pair $\langle\langle\mathbf{P}_{1,0,0,0},1\rangle;\{e\}\rangle$, which is in essence what a random variable is. Analogously, we describe it as $\langle\langle\mathbf{P}_{1,0,0,0},0\rangle;\{e\}\rangle$ when it is false, and  in this case it ``the particle $e$ does not possess the property $P_{1,0,0,0}$.'' In the original formulation of the KS contradiction, these properties where assigned to a single (same)  particle $\{e\}$. This requires us to make consistent valuations: if the property $P_{1,0,0,0}$ appears as $\mathbf{P}_{1,0,0,0}$ in different equations (that represent different contexts), its valuation must be the same, because, in the usual interpretation, it refers to \textit{the same particle} $\{e\}$. Let us denote by $V$ a variable assigning truth values (i.e., $V=0$ or $V=1$). 

But, if we use the resources of quasi-set theory, we cannot say that \textit{it is the same particle} which is possessing the  property $P_{i,j,k,l}$, but only that indistinguishable particles do possess it or not. We express this by an ordered pair as follows
\begin{equation}\label{Fed}
\langle\langle \mathbf{P}_{i,j,k,l};V\rangle;\llbracket x \rrbracket_z\rangle
\end{equation}
where $\llbracket x \rrbracket_z$ is a strong singleton of the class $[x]_z$ (defined by the $q$-object $x$ for a suitable qset $z$), and the pair of equation (\ref{Fed}) has q-cardinal 2. The above pair represents the proposition: ``there is one indistinguishable $\{x\}$ for which the property $P_{i,j,k,l}$ acquires the value $V$,'' but without any specific identification of the particle. Notice that, as said before, using the axioms of quasiset theory, it is formally impossible to identify the element of $\llbracket x \rrbracket_z$: we can only say that there is one of a kind (say, an electron). 

But now, if we try to make a concrete valuation using propositions represented by ordered pairs $\langle\langle \mathbf{P}_{i,j,k,l};V\rangle;\{x\}_{s}\rangle$, we realize that it is  consistent to assign different truth values to the same projection operator. Really, this can be done due to the fact that, if we consider two properties $P_{1}$ and $P_{2}$ represented by $\langle\langle\mathbf{P}_{i,j,k,l};1\rangle;\llbracket x \rrbracket_z\rangle$ and $\langle\langle \mathbf{P}_{i,j,k,l};0\rangle;\llbracket x' \rrbracket_z \rangle$, we cannot affirm (due to the fact that this is meaningless in a theory of truly indistinguishable entities) that $\llbracket x \rrbracket_z$ and $\llbracket x' \rrbracket_z$ are \textit{the same}. Thus, there is no contradiction in assigning the value $V=1$ to $\mathbf{P}_{i,j,k,l}$ in proposition $P_{1}$ and the value $0$ to proposition $P_{2}$. This modification of the propositional structure and the truth values assignment in the quantum formalism, allows us to avoid the contradiction in equations (\ref{eq:cabello-1st} and \ref{eq:cabello-1st-omega}). In other words, the description of propositions using quasiset theory allows us to assign definite values to particles, but in a way that is very compatible with the constrains imposed by the quantum formalism. Particles may have well-defined properties; we simply cannot tell which particle has which property.

\section{Conclusions and Final Remarks}\label{sec:Conclusions}

Let us write down these considerations in a more general form. Consider the set $\mathcal{B}(\mathcal{H})$ of bounded operators acting on a separable Hilbert space $\mathcal{H}$. It is well known that the collection $\mathcal{P}(\mathcal{H})$ of orthogonal projections is included in $\mathcal{B}(\mathcal{H})$ and that it forms an orthomodular lattice (which is modular for the finite dimensional case and strictly orthomodular in the infinite dimensional case). The KS theorem can be extended to more general von Neumann algebras (see \cite{doring_kochenspecker_2005}). Projection operators are interpreted as empirically testable propositions by appealing to the spectral theorem. Let $\mathbf{B}$ be the collection of Borel sets in $\mathbb{R}$ (the set of the reals). For each observable represented by a self-adjoint operator $A$, there exists a spectral measure 
$M_{A}:\mathbf{B}\longrightarrow\mathcal{P}(\mathcal{H})$ assigning to every Borel set $B$ a projection operator $M_{A}(B)$. The usual interpretation of $M_{A}(B)$ is the proposition ``the value of $A$ lies in $B$.'' But to be more precise, we should say what is the system for which this property is assigned. The above discussion indicates that the identity (or non-identity) of the system in question plays a crucial role. Let us call $Q$ our quantum system, and ask about the set theoretical nature of $Q$. Let us suppose first that $Q$ is identifiable and that it obeys the classical theory of identity. In this case, a more accurate description of $M_{A}(B)$ should be ``the value of $A$ lies in $B$ for the system $Q$.'' This can be naturally represented, in a standard set theoretical framework, as an ordered pair $\langle\langle M_{A}(B);1\rangle;\{q\}\rangle$, with the value ``$1$" meaning that the proposition is true. The negation of this proposition (i.e. ``the value of $A$ is not contained in $B$ for the system $Q$'') can be represented as $\langle\langle M_{A}(B);0\rangle;\{q\}\rangle$ (or equivalently, using the standard quantum logical interpretation of the orthogonal complement, we can write: $\langle\langle M_{A}(B);0\rangle;\{q\}\rangle$ ($(M_{A}(B))^{\bot};1\rangle;\{q\}\rangle=\langle\langle 1-M_{A}(B);1\rangle;\{q\}\rangle$). A valuation of the properties of $Q$ can be seen as a collection of ordered pairs of this form, each one of them having the truth value $V=1$ or $V=0$. These valuations must be consistent: we cannot have $\langle\langle P;1\rangle;\{q\}\rangle$ and $\langle\langle P;0\rangle;\{q\}\rangle$, because this would mean that $q$ possesses the empirically testable property defined by the projection operator $P$ and that it does not possesses it at the same time. This consistent valuation leads unavoidably to the KS contradiction, not only in $\mathcal{B}(\mathcal{H})$, but in more general von Neumann algebras as well.

If we now assume that quantum particles lack individuality, we must consider $Q$ as represented by an $q$-object $x$ in quasiset theory. This means that, the best we can do is to form a strong singleton for $q$: $\llbracket x \rrbracket_z$, taken from a suitable qset $z$. Thus, the propositional structure of quantum mechanics must be interpreted again. For the observable $A$ and the Borel set $B$, we can no longer claim that ``the value of $A$ lies in $B$ for the system $Q$.'' If $Q$ represents a truly non-individual entity (an entity devoid of identity), the correct way to make this assertion is to say that ``the value of $A$ lies in $B$ \textit{for one indistinguishable} of $Q$''. Thus, if we now want to represent this proposition using ordered pairs in our set-theoretical framework, we can write $\langle\langle M_{A}(B);1\rangle;\llbracket q \rrbracket_z\rangle$. But now, in the right hand side of the ordered pair, it  appears a strong singleton of the qset $[q]_z$, which stands for the indistinguishable from $q$ that belong to $z$ as seen above. This implies that potentially we have  a collection of indistinguishables from $\llbracket x \rrbracket_z$ that can be the value of the variable in the right hand side. In other words, any one of the $\llbracket x \rrbracket_z$'s can do the job. 

This of course gives us more freedom in claiming that a proposition is either true or false. And furthermore, now there is no contradiction between $\langle\langle M_{A}(B);1\rangle;\llbracket q \rrbracket_z\rangle$ and $\langle\langle M_{A}(B);0\rangle;\llbracket q \rrbracket_z\rangle$, simply because we cannot claim that the property pertains to a single and identifiable quantum system $Q$, but to any one of a collection of indistinguishable element of it. This gives us more freedom to choose valuations compatible with the constrains imposed by the quantum formalism and, thus, to avoid the KS contradiction.
 "The Ss
\medskip
In this paper we showed that it is possible to avoid  the KS contradiction using the theory of qsets to describe indistinguishable properties of particles. Our approach has some similarities to  that of Kurzynski \cite{kurzynski_contextuality_2017}, but here we not only provide a formal setup for avoiding the KS contradiction, but one that is mathematically precise and based on the ontology of quantum indistinguishability. It would be interesting to investigate the relationship between our approaches. 

Our results provide an example of how the use of non-standard mathematics may help us to solve conceptual problems in the interpretation of quantum mechanics. It also shows that taking a more critical look at the underlying ontological principles may lead to  interesting ways of thinking about some of the fundamental issues in quantum mechanics.  

\vspace{6pt} 

\acknowledgments{This research was partially conducted while JAB visited the Center for the Explanation of Consciousness, CSLI,  Stanford University, and he kindly thanks Prof. John Perry for his hospitality.  FH acknowledges CONICET and UNLP (Argentina) for partial financial support. DK  would like to thank the Brazilian council CNPq for partial support. The authors thank Pawel Kurzinsky, Ehtibar Dzhafarov, and Newton da Costa for enlightening discussions.}

\authorcontributions{All authors contributed equality in writing the paper.}

\conflictofinterests{The authors declare no conflict of interest.}

\bibliographystyle{mdpi}


\end{document}